\documentclass{an}

\usepackage{graphicx}
\usepackage{times}

\begin{document}

\Pagespan{1}{}
\Yearpublication{----}
\Yearsubmission{2015}
\Month{-}  
\Volume{---}  
\Issue{--}
\DOI{}

\title{Star formation rates and the kinematics of gas in the spiral 
       arms of NGC~628}

\author{A.\,S. Gusev\inst{1}\fnmsep\thanks{Corresponding author:
  \email{gusev@sai.msu.ru}},
F. Sakhibov\inst{2},
\and  Yu.\,N. Efremov\inst{1}
}

\titlerunning{Star formation and the kinematics of gas in NGC~628}

\authorrunning{A.\,S. Gusev, F. Sakhibov, \& Yu.\,N. Efremov}

\institute{
Sternberg Astronomical Institute, Lomonosov Moscow State University, 
Universitetsky pr. 13, 119992 Moscow, Russia
\and 
University of Applied Sciences of Mittelhessen, Campus Friedberg, 
Department of Mathematics, Natural Sciences and Data Processing, 
Wilhelm-Leuschner-Strasse 13, 61169 Friedberg, Germany}

\received{2015 Jan 20}
\accepted{2015 Apr 13}
\publonline{XXXX}


\abstract{Relations between star formation rates along the spiral arms 
and the velocities of gas inflow into the arms in grand-design galaxy 
NGC~628 were studied. We found that the radial distribution of average star 
formation rate in individual star formation regions in regular spiral arms 
correlates with the velocity of gas inflow into the spiral arms. Both 
distributions have maxima at a galactocentric distance of 4.5-5~kpc. There 
are no correlations between the radial distributions of average star formation 
rate in star formation regions in spiral arms and outside spiral arms in the 
main disc. We also did not find a correlation between the radial distribution 
of average star formation rate in star formation regions in spiral arms and 
H\,{\sc i} column density.}

\maketitle

\section{Introduction}

A connection between star formation and the spiral structure of disc 
galaxies is known since \nocite{morgan1953}Morgan, Whitford \& Code (1953) 
found concentrations of OB stars 
in the Sagittarius spiral arm of the Milky Way. This connection suggests 
that spiral arms trigger star formation. The development of the density-wave 
theory has shed new light on the connection between star formation and 
spiral arms. A density wave traveling through the disc of a galaxy 
produces a shock in the gas, which should trigger formation of stars in 
spiral arms. Spiral arms are dense and promote more gravitational 
instabilities and cloud collisions than the interarm regions, triggering 
molecular cloud formation and conglomeration in the arms. This can easily 
explain the observed concentration of young stars in spiral arms.

Many studies of the spiral structure in galaxies compare various observational 
data in order to link star formation with spiral density waves in 
discs. These comparisons are mostly qualitative. There is very illustrative 
diagram of \cite{roberts1975}, which relates the expected velocity of gas 
inflow into the spiral arm and the magnitude of the spiral arm shock with 
the morphological characteristics of the spiral structure and van den Bergh 
luminosity classes. The earlier the luminosity class of the galaxy (i.e., 
the higher its luminosity), the brighter and more conspicuous the spiral 
pattern it possesses. In this case, the galactic gas meets the spiral arm at 
a higher velocity. The higher this velocity, the stronger the shock 
of gas and the higher the luminosity of the galaxy.

A more detailed analysis one can make through the study of non-circular 
motions of gas clouds and stars caused by the spiral density wave in the 
disc of a galaxy. The perturbed velocities due to the spiral wave depend 
on the amplitude of the wave, the position of the corotation and Lindblad 
resonances, and the mass distribution in the galaxy (the shape of rotation 
curve). The same parameters affect the radial behaviour of the velocity of 
gas inflow into the spiral arm, and thereby probably also the radial profile 
of SFR in the disc.

Previously we compared a variation of the velocity of gas inflow, $v_\perp$, 
into the spirals and the surface density of star formation rate, 
$\Sigma_{\rm SFR}$, in the disc with the galactocentric distance $r$ in the 
spiral galaxy NGC~628 \nocite{sakhibov2004}(Sakhibov \& Smirnov 2004). We 
believe that the radial distribution of the SFR surface density, 
$\Sigma_{\rm SFR}$, over the disc has a ring form with the maximum at the 
galactocentric distance $r \sim 3$~kpc, whereas the maximum of the velocity 
of gas inflow $v_\perp$ is situated at $r \sim 5$~kpc. We derived the 
velocities of gas inflow $v_\perp$ into the spirals at different 
galactocentric distances via a Fourier analysis of the azimuthal distribution 
of the observed radial velocities in annular (ring form) zones of the disc 
of the NGC~628 and the radial profile of the surface density of SFR, using 
the empirical 'SFR vs. linear size' relation for star formation complexes 
\nocite{sakhibov2004}(Sakhibov \& Smirnov 2004) and coordinates, H$\alpha$ 
fluxes, and the sizes of H\,{\sc ii} regions presented by \cite{belley1992}.

\begin{table}
\caption[]{\label{table:param}
Basic parameters of NGC~628.
}
\begin{center}
\begin{tabular}{ll} \hline
Parameter                                & Value \\
\hline
Type                                     & SA(s)c \\
Total apparent $B$ magnitude ($B_t$)     & $9.70\pm0.26$ mag \\
Absolute $B$ magnitude ($M_B$)$^a$       & $-20.72$ mag \\
Inclination ($i$)                        & $7\degr\pm1\degr$ \\
Position angle (PA)                      & $25\degr$ \\
Heliocentric radial velocity ($v$)       & $659\pm1$ km\,s$^{-1}$ \\
Apparent corrected radius ($R_{25}$)$^b$ & $5.23\pm0.24$ arcmin \\
Apparent corrected radius ($R_{25}$)$^b$ & $10.96\pm0.51$ kpc \\
Distance ($d$)                           & 7.2 Mpc \\
Galactic absorption ($A(B)_{\rm Gal}$)   & 0.254 mag \\
Distance modulus ($m-M$)                 & 29.29 mag \\
\hline
\end{tabular}\\
\end{center}
\begin{flushleft}
{\footnotesize 
$^a$ Absolute magnitude of the galaxy corrected for Galactic extinction and
inclination effect. \\
$^b$ Isophotal radius (25 mag\,arcsec$^{-2}$ in the $B$-band) corrected for
Galactic extinction and absorption due to the inclination of NGC~628.}
\end{flushleft}
\end{table}

\begin{figure*}
\centering
\includegraphics[width=\textwidth]{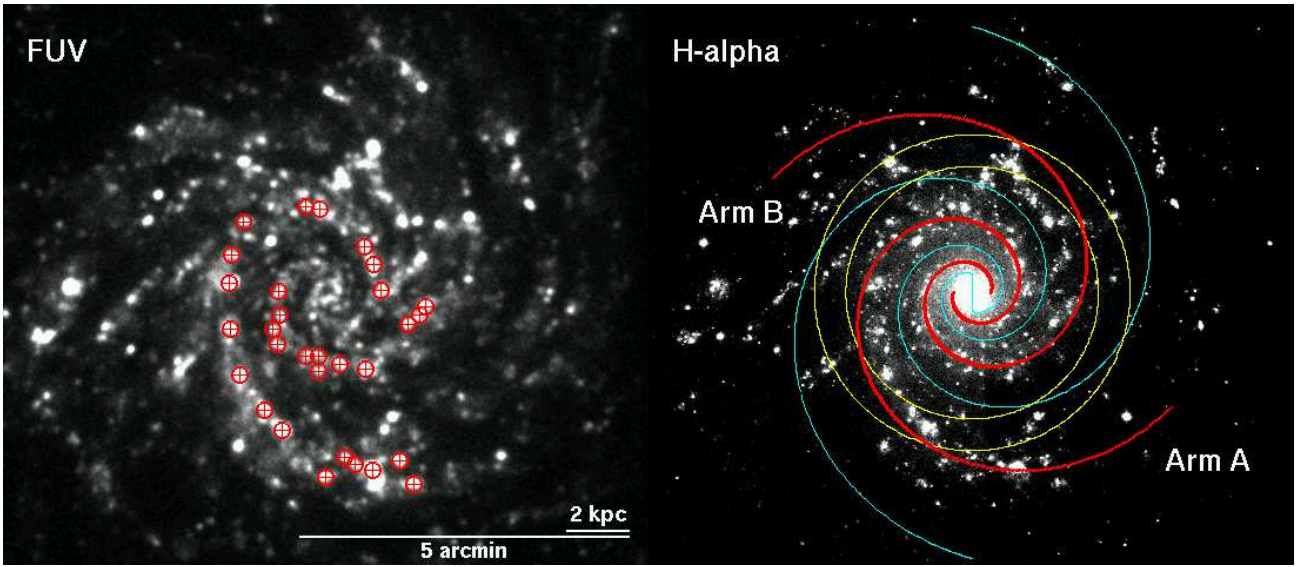}
\caption{(online colour at: www.an-journal.org) 
Images of NGC~628 in the FUV ({\it left}) and H$\alpha$ line ({\it right}). 
The FUV image was taken from {\it GALEX} archive, the H$\alpha$ image was 
obtained on the Maidanak Observatory with the 1.5~m telescope. Positions of 
star formation regions (red crosses in circles) from the list of 
\nocite{gusev2014} Gusev et al. (2014) are indicated in the FUV image. 
Logarithmic spiral arms (thick red curves) and the boundary between two parts 
of the galaxy, around Arm~A and Arm~B (thin cyan curve) are shown in the 
H$\alpha$ line image. Yellow circles in the H$\alpha$ image have radii 
$r = 4$ and 5 kpc (see the text for details). North is upward and east is 
to the left.
}
\label{figure:map}
\end{figure*}

The deviation between the maximum of the $\Sigma_{\rm SFR}$ in the disc at 
the $r \sim 3$~kpc and the maximum of the velocity of gas inflow $v_\perp$ 
at $r \sim 5$~kpc shows that the total galactic star formation rate in the 
main disc is not significantly enhanced by the presence of spiral arms (see 
\nocite{elmegreen2010}Elmegreen 2010).
 
But the question, whether the peak of the velocity of gas inflow $v_\perp$ 
into the spirals coincides with the peak of the radial distribution of the 
SFR within the spirals in the individual star formation regions, remains open.

Recently \cite{gusev2013} and \nocite{gusev2014}Gusev, Egorov \& Sakhibov 
(2014) investigated photometric 
properties of spiral arms in NGC~628, and location and parameters of star 
formation regions inside these arms. Thirty brightest star formation 
regions in the ultraviolet band, located in the spiral arms of NGC~628, were 
identified and studied (see Fig.~\ref{figure:map}). We found that the star 
formation regions in Arm~A are systematically brighter and larger 
than the regions in Arm~B \nocite{gusev2014}(Gusev et al. 2014). We also 
measured the star formation rates (SFRs) and the surface densities of SFR 
($\Sigma_{\rm SFR}$) within these star formation regions using obtained FUV 
magnitudes, H$\alpha$ luminosities and sizes. Since the star formation rate 
per unit area is higher in the arms than in the main disc we decide to study 
relations between SFRs along the arms and the velocities of gas inflow into 
the spiral arms.

The main goal of this new research is to study relations between SFRs along 
the arms and the velocities of gas inflow into the spiral arms. This study is
based on our own observations of the galaxy in H$\alpha$ line, obtained on 
the Maidanak Observatory with the 1.5~m telescope, {\it Galaxy Evolution 
Explorer (GALEX)} far-ultraviolet (FUV) data \nocite{gusev2013,gusev2014} 
(Gusev \& Efremov 2013; Gusev et al. 2014), as well as a Fourier analysis of 
the spatial distribution of the radial velocities of the neutral gas in the 
disc \nocite{sakhibov2004} (Sakhibov \& Smirnov 2004).

Observations, data reduction, determination of parameters of spiral arms, 
selection criteria and photometry of star formation regions have been 
described in detail in \nocite{gusev2013}Gusev \& Efremov (2013) and 
\nocite{gusev2014}Gusev et al. (2014).

\section{NGC~628, previous studies}

The spiral galaxy NGC~628 (M74) is one of the best studied nearby spiral 
galaxies viewed almost face-on (Fig.~\ref{figure:map}, 
Table~\ref{table:param}). This prototypical grand-design galaxy hosts two 
principal arms. The first (South arm by \nocite{rosales2011} 
Rosales-Ortega et al. 2011) is a long regular arm (Arm~A in 
Fig.~\ref{figure:map}). In this arm \cite{elmegreen1983} found a 
regular string of star complexes (H\,{\sc ii} regions). The shorter second 
arm (Arm B in Fig.~\ref{figure:map}) have the distorted outer part.
 
NGC~628 is an excellent example of a galaxy that has experienced recent star 
formation episodes. \cite{hodge1976} identified 730 H\,{\sc ii} regions in 
the galaxy. \cite{ivanov1992} observed and compiled a catalogue of positions, 
angular sizes, integral magnitudes and H\,{\sc ii} identification of 147 
OB associations and aggregates in NGC~628. Numerous star formation regions 
have been studied earlier based on photometric, spectroscopic and 
spectrophotometric observations 
\nocite{mccall1985,belley1992,ferguson1998,bresolin1999,larsen1999,
larsen2004,bruevich2007,rosales2011,gusev2012,berg2013} (McCall, Rybski \& 
Shields 1985; Belley \& Roy 1992; Ferguson, Gallagher \& Wyse 1998; 
Bresolin, Kennicutt \& Garnett 1999; Larsen 1999, 2004; Bruevich et al. 2007; 
Rosales-Ortega et al. 2011; Gusev et al. 2012; Berg et al. 2013).

Kinematics of the H\,{\sc i} disc has been studied by 
\nocite{shostak1984}Shostak \& van der Kruit (1984) and 
\nocite{kamphuis1992}Kamphuis \& Briggs (1992). The neutral hydrogen 
distribution correlates with 
the optical structure within the optical disc. The H\,{\sc i} velocity field 
out of the optical edge can be ascribed to circular rotation in a plane 
\nocite{shostak1984}(Shostak \& van der Kruit 1984). Two giant high velocity 
gas complexes ($M ({\rm H}$\,{\sc i}$)\sim(0.5-1)\times10^8M_\odot$) are 
located at $\sim10$~arcmin to the east and to the west from the galactic 
centre \nocite{kamphuis1992} (Kamphuis \& Briggs 1992). The gas velocity 
dispersion, as well as the stellar dispersion are typical for giant spiral 
galaxies, $\sigma({\rm H}$\,{\sc i}$) = 8-10$~km\,s$^{-1}$ at galactocentric 
distances, $r$, of $1.5-3.5$~arcmin (3.0--7.5~kpc for an adopted distance to 
NGC~628 of 7.2 Mpc; \nocite{shostak1984} Shostak \& van der Kruit 1984), 
$\sigma({\rm H}$\,{\sc ii}$) = 16-19$~km\,s$^{-1}$ at $r$ of 
$0.5-2.7$~arcmin (1.0--5.5~kpc) and $14\pm7$~km\,s$^{-1}$ at $r$ of 
$2.8-4.0$~arcmin (6.0--8.5~kpc; \nocite{fathi2007} Fathi et al. 2007), and 
$\sigma({\rm stars}) = 50\pm20$~km\,s$^{-1}$ at $r$ of $0.3-1.0$~arcmin 
(0.7--2.0~kpc; \nocite{macarthur2009} MacArthur, Gonz\'{a}lez \& Courteau 
2009).

Although NGC 628 is a member of a small group of galaxies \nocite{auld2006} 
(Auld et al. 2006), it cannot have undergone any encounter with satellites 
or other galaxies in the past 1 Gyr \nocite{wakker1991,kamphuis1992} 
(Wakker \& van Woerden 1991; Kamphuis \& Briggs 1992). 

The rotation speed of the arms (pattern speed) is determined along with 
other parameters of the spiral-density wave 
(\nocite{sakhibov2004}Sakhibov \& Smirnov 2004) via a Fourier analysis of 
the azimuthal distribution of the observed radial velocities 
(\nocite{shostak1984}Shostak \& van der Kruit 1984) in annular zones of the 
disc of NGC~628. Comparison of the pattern speed with the pure circular motion 
of the gas provided the velocity of gas inflow into the spiral arm at 
different radial distances. Corresponding corotation radius locates at 
$R_{\rm cor} \approx 7$~kpc (\nocite{sakhibov2004}Sakhibov \& Smirnov 2004).

The fundamental parameters of NGC~628 are presented in 
Table~\ref{table:param}. We take the distance to NGC~628, obtained in 
\cite{sharina1996} and \nocite{vandyk2006}van Dyk, Li \& Filippenko (2006). 
We used the position angle and the inclination of the galactic disc, derived 
by \nocite{sakhibov2004}Sakhibov \& Smirnov (2004).
The morphological type and the Galactic absorption, $A(B)_{\rm Gal}$, 
are taken from the NED data base.\footnote{http://ned.ipac.caltech.edu/} 
Other parameters are taken from the LEDA data 
base\footnote{http://leda.univ-lyon1.fr/} \nocite{paturel2003} (Paturel 
et al. 2003). We adopt the Hubble constant $H_0 = 75$ km\,s$^{-1}$Mpc$^{-1}$. 
With the assumed distance to NGC~628, we estimate a linear scale of 
34.9~pc\,arcsec$^{-1}$.

\section{Radial distributions of the star formation rates}

We derived the radial distributions of SFR in the galaxy 
using FUV and H$\alpha$ images of NGC~628. Previously, all data were 
corrected for Galactic absorption. Here we used the resulting ratio of the 
extinction in the {\it GALEX} FUV band to the color excess $E(B-V)$ 
$A_{\rm FUV}/E(B-V) = 8.24$ \nocite{wyder2007} (Wyder et al. 2007) and the 
standard ratio $c({\rm H}\alpha) = 0.308A(V)$.

Azimuthally averaged photometric profiles of the galaxy in FUV and H$\alpha$ 
were obtained using round apertures in steps of 2~arcsec. 
We do not transform images of NGC~628 to the face-on position, 
$i=0\degr$, because the position of any point in the galaxy is changed 
by value much less than the characteristic width of the spiral arm, if 
we use the adopted inclination $i=7\degr$. Difference between the uncorrected 
position and the position corrected for the inclination position angle 
is less than $0.5\degr$ and the galactocentric distance $\Delta r/r < 0.01$ 
for any point in the galaxy.

For estimation of SFR radial distribution peculiarities in the two opposite 
spiral arms of NGC~628, we divided the galactic image into two equal 
parts. The first part includes the area around Arm~A, and the second part -- 
around Arm~B. We built the boundary between the two parts using the 
parameters of symmetric logarithmic spirals obtained in \cite{gusev2013}. 
We determined the equation of the boundary at galactocentric distances 
$r\ge21.95$~arcsec as the equations of logarithmic spiral Arms~A and~B 
(see \nocite{gusev2013} Gusev \& Efremov 2013) that are rotated by $90\degr$ 
in the galactic plane. For the inner part of NGC~628, at 
$r<21.95$~arcsec, the boundary lies on the north-south line 
(Fig.~\ref{figure:map}).

We adopt the conversion factor of FUV luminosity to the star formation rate of 
\cite{iglesias2006}:
\begin{equation}
{\rm SFR (}M_\odot \,{\rm yr^{-1})} = 8.13\times10^{-44} L_{\rm FUV} {\rm (erg\,s^{-1})},
\label{equation:sfr_fuv}
\end{equation}
and the conversion factor of H$\alpha$ luminosity to star formation rate of
\cite{kennicutt1998}:
\begin{equation}
{\rm SFR (}M_\odot \,{\rm yr^{-1})} = 7.9\times10^{-42} L_{{\rm H}\alpha}
{\rm (erg\,s^{-1})}.
\label{equation:sfr_ha}
\end{equation}

To estimate physical values of SFR along the spiral arms, we introduce the term 
'SFR per unit of length of a spiral arm', $\Delta{\rm SFR}/\Delta l$. We 
adopt here, that all FUV and H$\alpha$ emission is concentrated in the thin 
spiral arms of the galaxy.

For a logarithmic spiral with a pitch angle $\mu$, the longitudinal 
displacement segment along the spiral, $\Delta l$, within a ringed aperture 
with a width $\Delta r$ is
\begin{equation}
\Delta l = \Delta r/\sin \mu
\label{equation:sp_long}
\end{equation}
independently of the galactocentric radius. 
\nocite{gusev2013}Gusev \& Efremov (2013) obtained a 
pitch angle $\mu = 15.7\degr$ for the arms. It is consistent with the 
estimations by \cite{danver1942} $\mu = 17\degr \pm 2\degr$ and close to the 
result of \nocite{kennicutt1976}Kennicutt \& Hodge (1976) who fitted every 
arm separately and obtained values $13.8\degr$ for Arm~A and $11.2\degr$ for 
Arm~B.

Thus, the values of $\Delta{\rm SFR}/\Delta l$ were obtained using
Eqs. (\ref{equation:sfr_fuv}), (\ref{equation:sfr_ha}) and 
(\ref{equation:sp_long}), and the constants $r=2$~arcsec (70~pc) and 
$\mu = 15.7\degr$.

\begin{figure*}
\vspace{7.0mm}
\centering
\includegraphics[width=\textwidth]{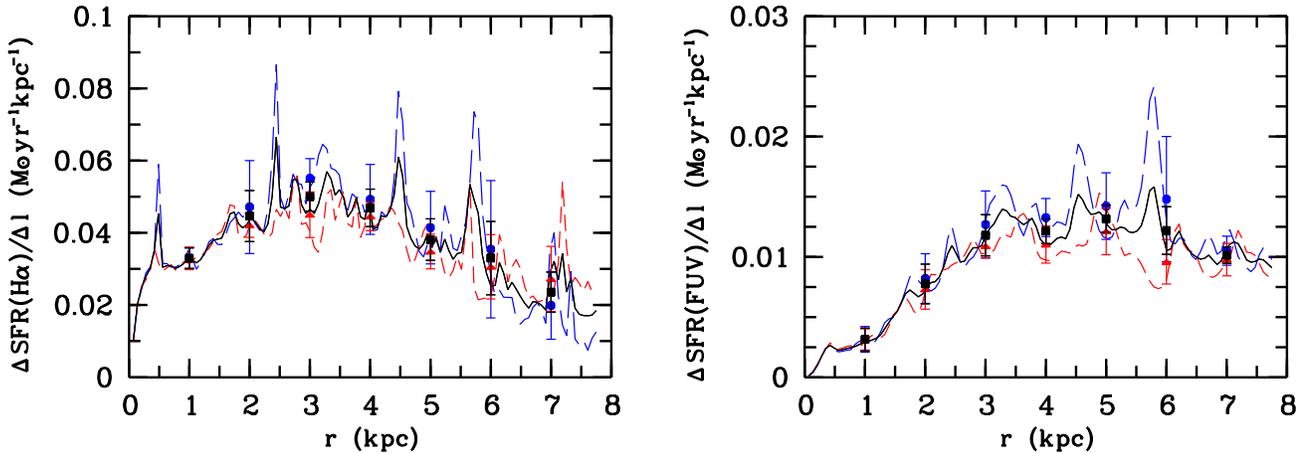}
\caption{(online colour at: www.an-journal.org) Radial distributions of 
SFRs per the unit of length of spiral arm, $\Delta{\rm SFR}/\Delta l$, 
based on the luminosities in H$\alpha$ ({\it left panel}) and FUV 
({\it right panel}), and their averaged values for Arm~A (blue 
long-dashed curves and circles), Arm~B (red short-dashed curves and 
triangles), and the mean for both arms (black solid curves and squares). 
The $\Delta{\rm SFR}/\Delta l$ averaged error bars are shown.
}
\label{figure:sfrp}
\end{figure*}

\begin{figure}
\vspace{7.0mm}
\centering
\includegraphics[width=0.47\textwidth]{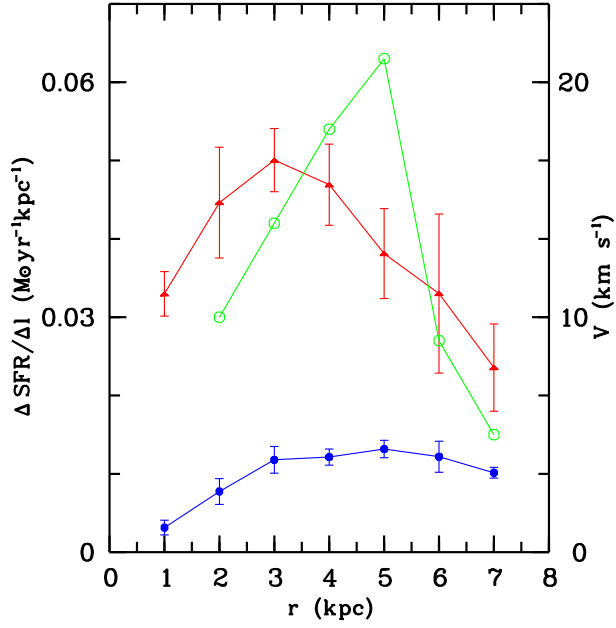}
\caption{(online colour at: www.an-journal.org) Radial distributions of 
averaged values of $\Delta{\rm SFR}/\Delta l$ based on the luminosities 
in H$\alpha$ (red line and triangles) and FUV (blue line and circles) 
for both arms. The $\Delta{\rm SFR}/\Delta l$ averaged error bars are shown. 
Radial distributions of the velocities of gas inflow into the spiral 
arms (green line and open circles) by \cite{sakhibov2004} are shown.
}
\label{figure:sfr}
\end{figure}

\begin{figure}
\vspace{7.0mm}
\centering
\includegraphics[width=0.47\textwidth]{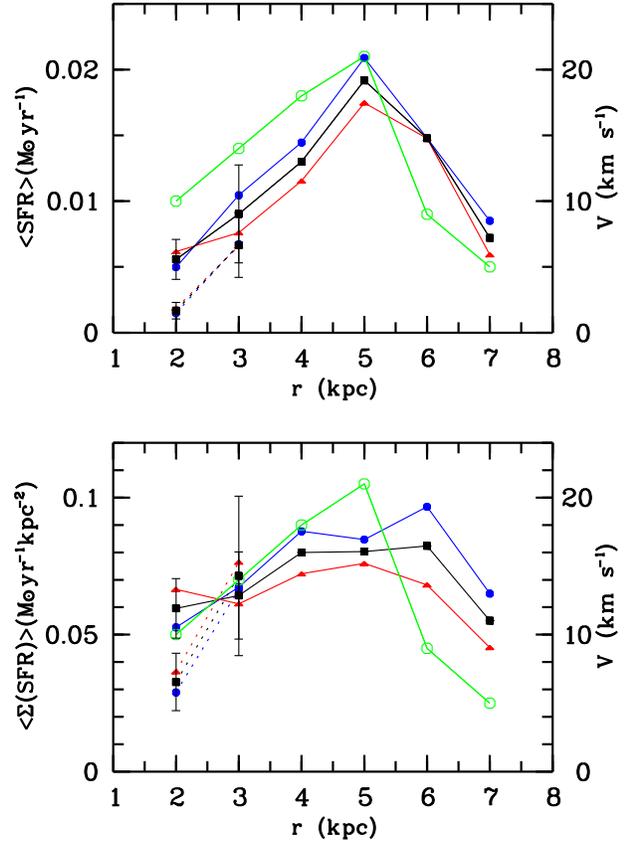}
\caption{(online colour at: www.an-journal.org) Radial distributions of 
the mean SFR ({\it top}) and the mean surface density of SFR ({\it bottom}) 
within star formation regions based on their luminosities in FUV (blue 
lines and circles), H$\alpha$ (red lines and triangles), and the mean 
for both luminosities (black lines and squares) in Arm~A (solid lines) 
and Arm~B (dotted lines). Radial distributions of the velocities of gas 
inflow into the spiral arms (green solid curves and open circles) are shown.
The $\langle{\rm SFR}\rangle$ and $\langle\Sigma_{\rm SFR}\rangle$ 
error bars are shown for Arms~A and B.
}
\label{figure:sfr2}
\end{figure}

The radial distributions of $\Delta{\rm SFR}/\Delta l$ are shown in 
Fig.~\ref{figure:sfrp} for Arms~A and B separately. We also give the mean 
radial distributions for both arms in this figure. Additionally, we 
calculated the values of $\Delta{\rm SFR}/\Delta l$, averaged over $r$ in 
steps of 1~kpc. These data are also presented in Fig.~\ref{figure:sfr}.

The radial distributions show slightly higher SFR in Arm A than in Arm~B at 
galactocentric distances from 2 to 6 kpc. However, the distributions based on 
FUV luminosities differ significantly from ones based on H$\alpha$ emissions. 
The distribution of $\Delta{\rm SFR (H}\alpha)/\Delta l$ has clear maximum 
at $r\approx3-4$~kpc, the distribution of $\Delta{\rm SFR (FUV)}/\Delta l$ 
has a flat plot at $r=3-6$~kpc (Fig.~\ref{figure:sfr}). 

The values of $\Delta{\rm SFR (H}\alpha)/\Delta l$ are 2--10 times larger 
than the values of $\Delta{\rm SFR (FUV)}/\Delta l$, this ratio decreases 
towards the outer part of NGC~628 (Fig.~\ref{figure:sfr}). These differences 
may be explained by the radial gradient of dust absorption in the disc of 
the galaxy. We roughly estimated parameters of dust distribution based
on the condition 
\begin{eqnarray}
{\rm SFR(FUV)/SFR(H}\alpha)(r)=1. \nonumber
\end{eqnarray}
The difference between FUV and 
H$\alpha$ luminosity distributions shown in Fig.~\ref{figure:sfr} can be 
explained by the gradient of dust absorption 
\begin{equation}
A(B)=1.545-0.134r,
\label{equation:abr}
\end{equation}
where $r$ is the galactocentric distance in units of kpc. If the dust is 
uniformly mixed in the stellar and gaseous medium, the dust absorption, 
$A(B)$, depends on the dust opacity, $\tau_B$, as 
\begin{equation}
A(B) = -2.5 \log{[1 -\exp(-\tau_B)]/\tau_B}
\label{equation:taur}
\end{equation}
(\nocite{disney1989}Disney, Davies \& Phillipps 1989). Using 
Eqs.~(\ref{equation:abr}) and (\ref{equation:taur}) we estimated the central 
opacity in the $B$ band, $\tau_B\approx3.9$, and the scale length of the dust 
disc, $h(B)_{\rm dust}\approx3.2$~kpc. The value of $h(B)_{\rm dust}$ is close 
to the disc scale length in the $B$ passband obtained by \cite{mollenhoff2004}, 
who estimated $h(B)=3.11$~kpc. The value of the central opacity in $B$ is 
typical for dust opacities in central regions of local universe 
spiral galaxies according to \cite{popescu2011}, who found $\tau_B\sim3.5-4$. 
We also estimated the total dust mass using Eq.~(44) from \cite{popescu2011}, 
$M_{\rm dust}\approx4\times10^7M_\odot$. It is close to the last estimation 
of the dust mass in NGC~628. \cite{aniano2012} found that 
$M_{\rm dust}=(2.9\pm0.4)\times10^7M_\odot$ using the data of {\it Spitzer} 
and {\it Hershel} (from 3.6~$\mu$m to 500~$\mu$m).

Nevertheless note that the radial distribution of the velocities of gas 
inflow into the spiral arms, $v_\perp$, poorly correlates with the 
distributions of SFRs based on both FUV and H$\alpha$ luminosities 
(Fig.~\ref{figure:sfr}).

\section{Star formation rates within star formation regions and the 
radial distribution of the velocity of gas inflow into the spiral arms}

To obtain the SFR, free from the dust influence, we used the data of SFR 
within star formation regions, which were obtained for 30 largest young 
stellar objects with known interstellar absorption in the regular part 
of spiral arms of NGC~628 in \nocite{gusev2014} Gusev et al. (2014). 
\nocite{gusev2014} Gusev et al. (2014) selected 30 star formation 
regions having a total magnitude corrected for interstellar absorption 
FUV$_0<$ 19.8 mag. Among regions fainter than 19.8~mag in the FUV, we found a 
large number of diffuse objects without strong H$\alpha$ emission. Moreover, 
there are no measurements of interstellar absorption for such faint objects. 
We divided the star formation regions into several groups depending on their 
distances to the galactic centre with a step of 1~kpc. For every group, the 
average SFR, $\langle$SFR$\rangle$, was calculated as 
\begin{equation}
\langle{\rm SFR}\rangle = \frac{1}{n} \sum_n {\rm SFR}_n,
\label{equation:sfr_mean}
\end{equation}
where $n$ is a number of objects in the 
group and SFR$_n$ is the SFR within the $n^{th}$ object. These mean SFRs 
were obtained separately for the objects in Arm~A and Arm~B based on both 
FUV and H$\alpha$ data (Fig.~\ref{figure:sfr2}). We also calculated the mean 
SFRs for both FUV and H$\alpha$ luminosities as 
($\langle$SFR(FUV)$\rangle$+$\langle$SFR(H$\alpha)\rangle$)/2.

Note that the total SFR within studied 30 regions, 
$\approx 0.25~M_\odot \,{\rm yr^{-1}}$, is a significant part (one 
third) of the total SFR in NGC~628, $0.7\pm0.2~M_\odot \,{\rm yr^{-1}}$ 
(\nocite{calzetti2010}Calzetti et al. 2010).

Similarly we calculated the mean values of SFR surface density 
within the star formation regions, $\langle\Sigma_{\rm SFR}\rangle$. They 
are measured as
\begin{equation}
\langle\Sigma_{\rm SFR}\rangle = \frac{4}{\pi}\langle\frac{\rm SFR}{d^2}\rangle,
\label{equation:ssfr_mean}
\end{equation}
where diameters of star formation regions, $d$, were obtained in 
\nocite{gusev2014} Gusev et al. (2014).

Arm~B is distorted at the galactocentric distances $r>3.2$~kpc 
(Fig.~\ref{figure:map}). We did not measure the luminosities of star 
formation regions in the distorted part of Arm~B. As a result, the radial 
distributions of $\langle{\rm SFR}\rangle$ and 
$\langle\Sigma_{\rm SFR}\rangle$ within the star formation regions 
in Arm~B were constructed up to $r\approx3$~kpc only.

As seen from Fig.~\ref{figure:sfr2}, the values and the distributions of 
$\langle{\rm SFR}\rangle$ and $\langle\Sigma_{\rm SFR}\rangle$ obtained from 
FUV and H$\alpha$ data are close to each other. The mean SFR within star 
formation regions in Arm~A is comparable within the errors with the 
mean SFR within star formation regions in Arm~B.

Distribution of the mean SFR within star formation regions in Arm~A 
correlates with the distribution of the velocities of gas inflow into the 
spiral arm very well. Both distributions have clear maximum at $r=5$~kpc. 
The mean surface density of SFR within star formation regions does not 
correlate with $v_\perp$, it is approximately constant at $r=3-6$~kpc, 
$\langle\Sigma_{\rm SFR}\rangle = 0.06-0.09$~Myr$^{-1}$kpc$^{-2}$, and it 
falls down towards corotation region of NGC~628 (Fig.~\ref{figure:sfr2}).

\begin{figure}
\vspace{7.0mm}
\centering
\includegraphics[width=0.44\textwidth]{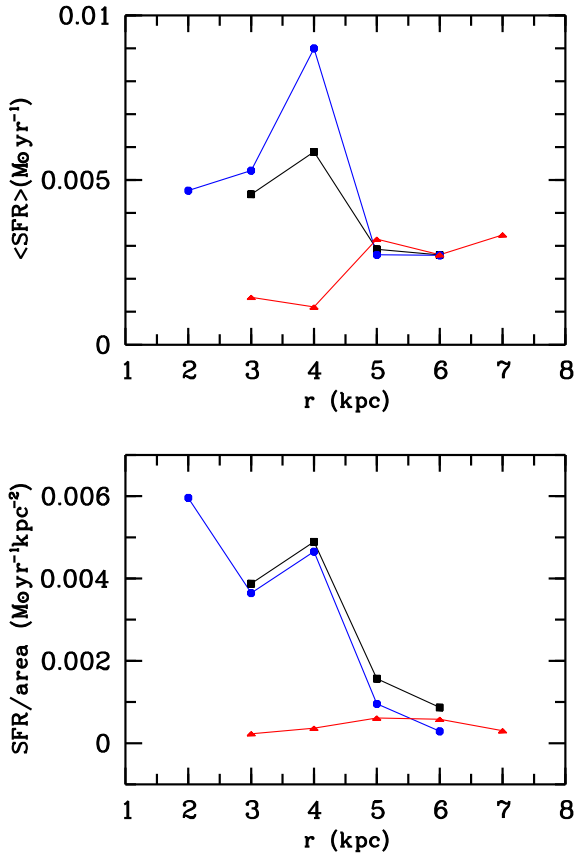}
\caption{(online colour at: www.an-journal.org) Radial distributions of the 
mean SFR ({\it top}) and the total SFR per square kpc ({\it bottom}) within 
star formation regions based on their luminosities in H$\alpha$ by 
\cite{rosales2011}. Distributions of SFR within star formation regions in 
spiral arms (blue lines and circles), within star formation regions outside 
spiral arms in the main disc (red lines and triangles), and within all star 
formation regions (black lines and squares) are shown.
}
\label{figure:sfr3}
\end{figure}

Difference between the radial distributions of $\langle{\rm SFR}\rangle$ 
and $\langle\Sigma_{\rm SFR}\rangle$ is a result of the fact that the star 
formation regions in Arm~A on the galactocentric distances 4~kpc~$<r<$~6~kpc 
are larger and brighter than those in the inner and outer parts of the arm.

A small number of objects in our sample, containing only bright 
star formation regions, makes it impossible to determine the maximum in 
the radial distribution of SFR with an accuracy less than $\pm1$~kpc. 
That is why we additionally examined a sample of H\,{\sc ii}~regions from 
\nocite{rosales2011}Rosales-Ortega et al. (2011) containing 96 objects 
inside and outside the spiral arms of the galaxy. The results of 
the analysis of this sample are presented in Fig.~\ref{figure:sfr3}. Here, 
we calculated the average SFR using Eq.~(\ref{equation:sfr_mean}). In 
contrast to the bottom panel of Fig.~\ref{figure:sfr2}, where we plotted the 
profile of the mean SFR within star formation regions per area occupied by 
a star formation region, calculated by Eq.~(\ref{equation:ssfr_mean}), in the 
bottom panel of Fig.~\ref{figure:sfr3} we present the profile of the total 
SFR within star formation regions per the galaxy area, averaged by $r$,
\begin{eqnarray}
{\rm SFR/area} = \sum_n {\rm SFR}_n/\pi[(r+0.5)^2-(r-0.5)^2], \nonumber
\end{eqnarray}
where $r$ is a galactocentric distance in units of kpc.

Only two objects in the spiral arms have $r=7\pm0.5$~kpc and two star 
formation regions outside the spiral arms are located at distance 
$r=2\pm0.5$~kpc. We do not show data for them in Fig.~\ref{figure:sfr3}.

As seen from Figs.~\ref{figure:sfr2} and \ref{figure:sfr3}, the maximum in 
distribution of the mean SFR within star formation regions in spiral arms by 
\cite{rosales2011} is shifted relative to the maximum in distribution of the 
mean SFR within star formation regions in Arm~A by our data (4~kpc versus 
5~kpc). The basic reason of this shift is a presence of several bright 
H\,{\sc ii}~complexes, which are inside Arm~B in the northern-western 
part of NGC~628; these complexes are located at $r\approx4-4.5$~kpc 
(Fig.~\ref{figure:map}). The largest and brightest in H$\alpha$ complex 
No.~40 by \cite{rosales2011} is also located here. Moreover, a significant 
contribution to the increase in $\langle{\rm SFR}\rangle$ is made by one of 
the brightest star formation regions, No.~A12 
(\nocite{gusev2014}Gusev et al. 2014) = 84 
(\nocite{rosales2011}Rosales-Ortega et al. 2011), which is located at 
$r\approx4.5$~kpc. The estimated galactocentric distance to his centre is 
4.60~kpc, however \cite{rosales2011} gave the distance $r=4.43$~kpc.

The shift in absolute values of the mean SFR between the distributions 
based on our data and the data of 
\nocite{rosales2011}Rosales-Ortega et al. (2011) is due to the fact that 
the sample of \nocite{rosales2011}Rosales-Ortega et al. (2011) contains more 
H\,{\sc ii}~regions, than ours, and includes fainter regions not included 
in our sample.

Note that the maximal velocity of gas inflow into the spiral arms is 
apparently achieved at distances $r\sim4.5-5.0$~kpc, as can be seen from 
the velocity graph in Fig.~\ref{figure:sfr2}.

Total SFR within star formation regions in spiral arms per the galaxy area 
decreases considerably with increasing distance from the centre of the 
galaxy (bottom panel of Fig.~\ref{figure:sfr3}). Nevertheless, the increase 
in star formation rate at $r=4$~kpc seen clearly in the figure.

Radial distributions of the mean and total SFRs within star formation 
regions inside and outside the spiral arms are significantly {\rm different} 
each from other (Fig.~\ref{figure:sfr3}). Both $\langle{\rm SFR}\rangle$ and 
total SFR per galaxy area within star formation regions in spiral arms have 
obvious maxima at $r\approx4$~kpc and they decrease towards 
$r\sim6$~kpc. $\langle{\rm SFR}\rangle$ and the total SFR within star 
formation regions outside the spiral arms grow on the distances from 4 to 
6~kpc (Fig.~\ref{figure:sfr3}). Both $\langle{\rm SFR}\rangle$ and total 
SFR per galaxy area within star formation regions become equal to each other 
for the objects inside and outside the spiral arms at $r\approx6$~kpc.

\begin{figure}
\vspace{7.0mm}
\centering
\includegraphics[width=0.47\textwidth]{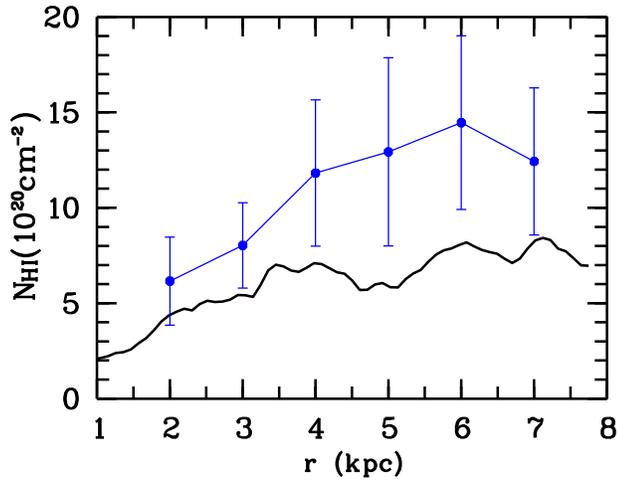}
\caption{(online colour at: www.an-journal.org) Radial distributions of 
H\,{\sc i} column density (thick black curve) and the mean H\,{\sc i} column 
density in H\,{\sc i} patches in spiral arms associated with 
photodissociation regions (blue lines and circles). The 
$N_{\rm HI}$ averaged error bars are shown.
}
\label{figure:hi}
\end{figure}

Obviously, star formation parameters depend on the pro-perties of 
interstellar medium. The key ingredient of star formation is a molecular 
hydrogen. However, molecular hydrogen is concentrated in the central part of 
galaxies. In contrast to H$_2$, a density of neutral hydrogen decreases 
towards the centre (deficiency of H\,{\sc i} is typically observed at the 
distance of a few kpc from the centre). We examined the radial distribution 
of H\,{\sc i} column density in NGC~628 based on the 21~cm line image 
obtained in \cite{walter2008}. The image\footnote{Original image: 
NGC\_628\_RO\_MOM0:I:HI:wbb2008.fits} was downloaded from the NED data 
base (Fig.~\ref{figure:hi}). \cite{heiner2013} estimated atomic hydrogen 
column densities in H\,{\sc i} patches in the spiral arms associated with 
photodissociation regions in NGC~628. Radial distribution of the mean 
H\,{\sc i} column density in these H\,{\sc i} patches, averaged by $r$, is 
also presented in Fig.~\ref{figure:hi}.

As seen from Fig.~\ref{figure:hi}, typical $N_{\rm HI}$ values in H\,{\sc i} 
patches are $1.5-2$ times larger than the values averaged azimuthally. 
However, their profiles are similar to each other; the H\,{\sc i} column 
densities increase up to the galactocentric distance $3-4$~kpc and 
they have plateau-like profiles at $r=4-7$~kpc. These profiles do not 
coincide with the profiles of the radial distribution of 
$\langle{\rm SFR}\rangle$ and total SFR per galaxy area within star formation 
regions in spiral arms of the galaxy (Figs.~\ref{figure:sfr3}, 
\ref{figure:hi}).

\section{Discussion}

We would like to underline the important result derived in this paper: the 
peak of the orthogonal component of the velocity of gas inflow into the 
spiral arms, $v_\perp$, coincides with the peak of the radial distribution 
of the mean SFR within the individual star formation regions. Both peaks 
are placed within the regular spiral arms at the galactocentric distance 
$r\approx4.5-5.0$~kpc, where the component of the difference between the 
spiral pattern and gas velocities, that is orthogonal to the arm, is maximal.

This result confirms the prediction of the density wave theory 
illustrated in \nocite{roberts1975} Roberts et al. (1975). 
\nocite{sakhibov2004}Sakhibov \& Smir-nov (2004) showed similar radial 
distributions of the velocity of gas inflow and the surface density of SFR 
in another nearby spiral galaxy NGC~6946. The maximum of the velocity 
of gas inflow and the peak of SFR are at the same radial distance 
$r\approx4$~kpc from the centre of the galaxy, between the 
inner Lindblad resonance ($\sim2.5$~kpc) and the corotation radius 
($\sim9$~kpc) (\nocite{sakhibov2004d}Sakhibov 2004).

Similar situation was noted in the SW arm of M31, whe-re the stellar age 
gradient across this arm segment is clearly observed, being evident due 
to the unusually large pitch angle of this arm segment 
(\nocite{efremov2010}Efremov 2010), whereas the pitch angle is close to zero 
for almost all other arms of M31. According to the classical theory 
(\nocite{roberts1975}Roberts et al. 1975), the degree of gas compression by 
a spiral shock wave is determined by the component $v_\perp$ of difference 
between the velocities of solid-body rotation of the density wave ($Vdw$) 
and differential rotation of the galaxy's gas, $V$, around its centre. The 
component $v_\perp$ is perpendicular to the wave front, i.e. to the inner 
boundary of the arm: $v_\perp = (V - Vdw) \sin \mu$ (see 
\nocite{efremov2010}Efremov 2010).

It implies that at the same distance from the co-rotation (where $V = Vdw$), 
the shock wave should be stronger in the arm segment with the larger pitch 
angle. The shock wave compresses gas clouds; the higher the density of the 
initial gas cloud, the higher the effectiveness of star formation. The 
large pitch angle of the arm segment in question leads to a high value of 
$v_\perp$ (see Fig.~\ref{figure:map}).

Peaks in the radial distribution of the gas inflow velocity into the spiral 
arms, as well as in the mean SFR within star formation regions in spiral arms 
do not coincide with the peak of the radial distribution of H$\alpha$ 
luminosity in the galaxy at $r\approx3$~kpc. Apparently, this difference may 
be explained by the radial gradient of dust absorption in the disc of NGC~628 
as well as diffuse radiation in H$\alpha$ in the central part of the galaxy.

On the other hand, the radial distributions of the mean and total SFRs in the 
individual star formation regions inside and outside the spiral arms 
significantly differ from each other (Fig.~\ref{figure:sfr3}). This result 
confirms the conclusion made previously by \cite{elmegreen2010}, that the 
total galactic star formation rate in the main disc is not significantly 
enhanced by the presence of spiral arms.

The mean surface density of SFR within star formation regions is approximately 
constant at the distances from 3 to 6~kpc from the centre of NGC~628 
(Fig.~\ref{figure:sfr2}). Obviously, the mean surface density of SFR within 
star formation regions, $\Sigma_{\rm SFR}$, depends basically on general 
properties of interstellar medium such as gas density and pressure.

Both the mean and the total SFRs within star formation regions become equal to 
each other for the objects inside and outside the spiral arms in the outer 
part of NGC~628 at $r>5$~kpc (Fig.~\ref{figure:sfr3}). These regions are 
situated in the vicinity of the corotation radius, $R_{\rm cor}\approx7$~kpc, 
where the velocity of the spiral pattern coincides with the rotation velocity 
of the galaxy. Dynamic properties of interstellar matter inside and outside 
the spiral arms must be close to each other in this region.

We note in conclusion, that as 
\nocite{martinez2014}Mart\'{i}nez-Garc\'{i}a \& Puerari (2014) have noted, 
most of a dozen galaxies without signatures of density-formed spiral pattern 
(\nocite{foyle2011}Foyle et al. 2011) have spiral arms of the flocculent 
and multi-arm type (apart from NGC~628 and NGC~5194). These galaxies are not 
described by the density wave theory.

\section{Conclusions}

We found correlation between the mean SFR within individual star formation 
regions in the regular parts of spiral arms and the velocity of gas inflow 
into the spiral arms in the grand-design galaxy NGC~628. Both quantities reach 
a maximum at a distance of 4.5--5~kpc from the centre of the galaxy.

Radial distributions of SFRs within star formation regions inside and 
outside the spiral arms are not correlated with each other. However, both the 
mean and the total SFRs within star formation regions become approximately 
equal to each other for the objects inside and outside the spiral arms in the 
outer part of NGC~628 on the distance $r\sim6$~kpc, near the corotation 
radius.

The mean surface density of SFR within star formation regions is approximately 
constant at the distances from 3 to 6~kpc from the centre of NGC~628.

\acknowledgements
 We are grateful to the referee for his/her constructive comments. 
 The authors thank A.V.~Zasov (SAI MSU) for helpful discussions and 
 E.V.~Shimanovskaya (SAI MSU) for help with the editing of this paper. 
 The authors acknowledge the usage of the HyperLeda data base 
 (http://leda.univ-lyon1.fr), the NASA/IPAC Extragalactic Database 
 (http://ned.ipac.caltech.edu), B.A. Miculski archive for space 
 telescopes (http://galex.stsci.edu), and the Padova group online server 
 CMD (http://stev.oapd.inaf.it). This study was supported by the 
 Russian Science Foundation (pro-ject no. 14--22--00041).

\end{document}